\documentclass[aps,prd]{revtex4}
\usepackage{graphicx}
\newcommand\bef{\begin{figure}}
\newcommand\eef[1]{\label{fg:#1}\end{figure}}
\newcommand\beq{\begin{equation}}
\newcommand\eeq[1]{\label{#1}\end{equation}}
\newcommand\beqa{\begin{eqnarray}}
\newcommand\eeqa[1]{\label{#1}\end{eqnarray}}
\newcommand\bet{\begin{table}}
\newcommand\eet[1]{\label{tb:#1}\end{table}}

\newcommand\fgn[1]{Figure \ref{fg:#1}}
\newcommand\eqn[1]{eq.\ (\ref{#1})}

\newcommand\tbn[1]{Table \ref{tb:#1}}

\newcommand{\C}{{\cal C}}
\newcommand\cl[1]{#1\%{\rm\ CrI\/}}
\newcommand{\cfr}{{\rm CFR\/}}

\newcommand\erfi{{\rm Erfi\/}}

\newcommand\etal{{\sl et al.\/}}
\newcommand\ie{{\sl i.e.\/}}
\newcommand{\I}{{\cal I\/}}

\begin{document}

\author{Sourendu Gupta}
\affiliation{Department of Theoretical Physics, Tata Institute of Fundamental
 Research,\\ Homi Bhabha Road, Mumbai 400005, India.}
\title{Epidemic parameters for COVID-19 in several regions of India}
\begin{abstract}
Bayesian analysis of publicly available time series of cases and
fatalities in different geographical regions of India during April
2020 is reported.  It is found that the initial apparent rapid growth
in infections could be partly due to confounding factors such as
initial rapid ramp-up of disease surveillance. A brief discussion is
given of the fallacies which arise if this possibility is neglected.
The growth after April 10 is consistent with a time independent but
region dependent exponential. From this, $R_0$ is extracted using both
known cases and fatalities. The two estimates are seen to agree in many
cases; for these {\cfr} is reported. It is seen that {\cfr} and $R_0$
increase together.  Some public health implications of this observation
are discussed, including a target doubling interval if medical facilities
are to remain adequate.
\\ \bigskip 
TIFR/TH/20-16\\ 
\end{abstract}
\maketitle

\section{Introduction}

SARS-COV-2 is a virus which has newly entered the global human population
\cite{when}. As this host-parasite system evolves towards an equilibrium,
its epidemiology has been studied extensively, but with some conflicting
results \cite{conflict1,conflict2}. The true extent of its penetration
into the population is as yet open to question \cite{serotesting}, since
testing is fairly restricted in most countries. Nor is the progression
of the disease, COVID-19, or its method of spread completely clear
\cite{review}. Since the virus is already so widely established, it seems
unlikely that it will be totally eliminated soon. So it is important to
extract basic epidemiological parameters as cleanly as possible.

India has managed to geographically contain the spread of the COVID-19
epidemic with the nation-wide lock-down which started on 24 March, 2020.
At the end of April the proportion of identified cases in India as a whole
was a few tens per million, with 1--2 orders of magnitude more in hot
spots.  Even if this were wrong by an order of magnitude, it would still
mean that the epidemic remains at an early stage in India. This, combined
with the lock-down, presents an opportunity to examine the growth of the
epidemic in multiple isolated regions which implement essentially the
same policy with regard to testing. This study examines the heterogeneity
in the growth rate of the disease, in several ways. First, the doubling
intervals, $\tau$, of the cumulative number of identified cases, $C(t)$,
and the cumulative number of fatalities, $D(t)$, is examined. From $\tau$
it is possible to extract the basic reproduction rate, $R_0$, within
epidemic models. Marked heterogeneity are observed. After this the
correlation between the case fatality ratio, \cfr, and $R_0$ is studied.

Epidemic data, especially at the beginning is never clean. The public
health system has to gear up for disease surveillance. The continuing
recurrence of Cholera epidemics \cite{cholera}, the spread of Dengue
and Chikungunya \cite{chikungunya}, the successful surveillance and
elimination of Nipah \cite{nipah} and Zika \cite{zika}, show that
India has a mixed record on epidemic surveillance.  In addition to a
possible lag between the beginning of the epidemic and its surveillance,
there could be a problem of incomplete surveillance during the time the
health service ramps up. Any examination of data has to allow for the
identification of confounding factors such as these.

For the COVID-19 surveillance data, there are further cautionary remarks.
The ICMR guidelines for testing \cite{icmr} specify that only symptomatic
cases should be tested using rRT-PCR. This part of the policy has been
unchanged since the middle of March. Depending on the fraction of cases
which are symptomatic, this could miss the actual prevalence of the
disease in the population. Estimates of the fraction of asymptomatic
infections range as high as 80\% \cite{cebm}, implying that, in this
extreme case, the testing policy can never reveal more than 20\% of the
cases. The social stigma attached to COVID-19 \cite{stigma} also means
that some fraction of infections may be cryptic.

There are uncertainties in the statistics of fatalities also. It has
been reported that in Europe and the US the number of fatalities due to
COVID-19 may have been underestimated by a factor of 2--3. Indian cities
have fairly complete registries of deaths, so miscounting of COVID-19
fatalities could come mainly from mistaken or incomplete reports of
the cause of death. For larger regions, say districts and whole states,
where most deaths happen at home and death certificates are not common
\cite{million} the errors in counting fatalities may be significantly
larger, and hard to estimate at this time.

One point about the quality test that is developed here is that absolute
numbers are not as important for it as the check that fatalities and
identified cases are independently tracing the same rate of growth of
the epidemic. This is expected at the beginning of the epidemic, when
all epidemic models become linear, and the growth of generic measures
is driven by the maximum eigenvalue of the linearized models. However,
in the extraction of the \cfr, the absolute counts do matter.  In spite
of the uncertainties, the correlation of {\cfr} and $R_0$ holds important
lessons for public health in the inevitable later stages of this epidemic
in India, and the middle and low income countries of the world.

\section{Method}

\subsection{Data}

Data has been extracted from official sources where possible. For
Ahmedabad city, the data is made publicly available by the municipal
council of the city \cite{ahmedabad}. This well-organized site corrects
data retrospectively for up to about 10 days.  For Chennai city, the
data has been collected from daily tweets by the municipal council
\cite{chennai}.  For Indore city, the data was collected from daily
bulletins of the Chief Medical and Health Officer and the collection is
available for public use \cite{indore}.  For Mumbai city the data has
been collated from the daily tweets by MCGM \cite{mcgm} into a publicly
available site \cite{prahlad}. For Pune district the data was collated
from the daily tweets by the district authority \cite{pune} and the
collection is publicly available \cite{indu}.  For Delhi and all other
states, the data was taken from the publicly available collection at
Covid19India \cite{covid19india}. This site corrects data retrospectively
for over a week. Only data on the cumulative number of identified cases,
$C(t)$, and the cumulative number of known fatalities, $D(t)$, are used
in this analysis. For this work data collection stopped on May 1, 2020,
and retrospective corrections made after this are not included.

The unquantifiable parts of the errors in the counts of cases and
fatalities due to COVID-19 were discussed in the previous section,
along with the reasons why their estimates need not be included in this
analysis. However, there is another part of the errors in the daily counts
of cases and fatalities which come from backlogs of tests or hospital
records. These shuffle a fraction of the numbers from one day to another,
and therefore cause errors in the daily counts. As long as the number
of facilities keep pace with the growth of the epidemic, these errors
remain proportional to the number of cases and fatalities. Since the wait
time for hospital beds for COVID-19 cases has remained roughly constant
during the period of study, this argument is expected to hold. In view of
this, of 20\% of the reported values of $C(t)$ and $D(t)$ are assigned
as errors. The specific fraction, 20\%, was chosen to in order to cover
the long range fluctuations visible in the time series (for example in
those visible on days 3 and 13 of \fgn{timedep}). It has been seen that
official reports and independent estimates of these number are generally
within this range.

\subsection{The doubling interval and $R_0$}

At such an early stage in the infection, it is reasonable to
assume exponential growth, \ie, doubling every $\tau$ days. Within this
assumption one can check how well the lock-down is working by letting
the doubling interval become time dependent.  The simplest function to
try is a linear change in $\tau$, \ie,
\beq
   C(t) = \C_0 2^{t/(\tau_0+t\tau_1)},
\eeq{timedependent}
and a similar set of three parameters for $C(t)$.  Note that $\tau_0$
has dimensions of time, whereas $\tau_1$ is dimensionless. A fitting form
with constant doubling interval was also used; this is denoted $\tau$,
dropping the subscript.

The fitting procedure follows the methods of \cite{previous}, with gamma
distributions used as prior probability distribution functions (PDFs)
for $\tau_0$ and $\C_0$. The additional parameter $\tau_1$ is allowed
to take positive and negative values, by letting the prior PDF to be a
Gaussian. For all these distributions, the widths are taken large enough
that the posterior distribution is insensitive to the choice of priors.

The appendix contains details of the relation between a time varying
doubling interval and time variation of the basic reproductive rate
$R_0$. This requires choosing a model of the epidemic. Using the SEIR
model, and the median interval between the appearance of symptoms and
the time of fatalities, $t_2=17.8$ days \cite{lancet}, one has
\beq
 R_0^0 = 1+\frac{t_2\ln2}{\tau_0}, \qquad{\rm and}\qquad
 \frac{R_0^1}T = -\frac{2t_2\tau_1\ln2}{\tau_0^2}.
\eeq{rematch}
When a constant $\tau$ is used, one can set $\tau_1=0$ in the above
formul{\ae} and write $R_0$ and $\tau$ instead of $R_0^0$ and $\tau_0$.

Exactly the same procedure is followed for fits to the time series for
$D(t)$. Estimates of the median values of the parameters, along with
interquartile ranges (IQR) and 95\% credible intervals (CrI) are quoted
for the doubling intervals as well as $R_0$.

\subsection{The case fatality ratio}

The analysis of the time series for $C(t)$ and $D(t)$ lead quite naturally
to the case fatality ratio, \cfr. This is defined as the ratio
\beq
 \cfr = D(t)/C(t).
\eeq{cfr}
If $C(t)$ is underestimated, then {\cfr} is overestimated, and conversely,
when $D(t)$ is underestimated, then {\cfr} is also underestimated. This
was regulated using a Bayesian estimator. Since the outcome is binomial,
the prior PDF used is a beta distribution
\beq
  P(\cfr) = \frac1{B(\alpha,\beta)}\;\cfr^{\alpha-1}(1-\cfr)^{\beta-1},
\eeq{betadist}
with $\alpha=1$ and $\beta=2$. These choices make the posterior
distribution insensitive to doubling or halving the values of
the priors. The posterior distribution is of the same form with
$\alpha=1+D(t)$ and $\beta=2+C(t)-D(t)$, with $t$ taken to be the final
day of the analysis. Since $C(t)$ and $D(t)$ are both large, the following
approximations for the median, $\mu$, and standard deviation, $\sigma$,
may be used:
\beq
 \mu[\cfr] \approx \frac{\alpha-1/3}{\alpha+\beta-2/3},
   \qquad{\rm and}\qquad
 \sigma^2[\cfr]  = \frac{\alpha\beta}{(\alpha+\beta)^2(\alpha+\beta+1)}\,.
\eeq{cfrestimators}

\section{Results}

\subsection{On $R_0$}

\bef
\begin{center}
\includegraphics[scale=0.5]{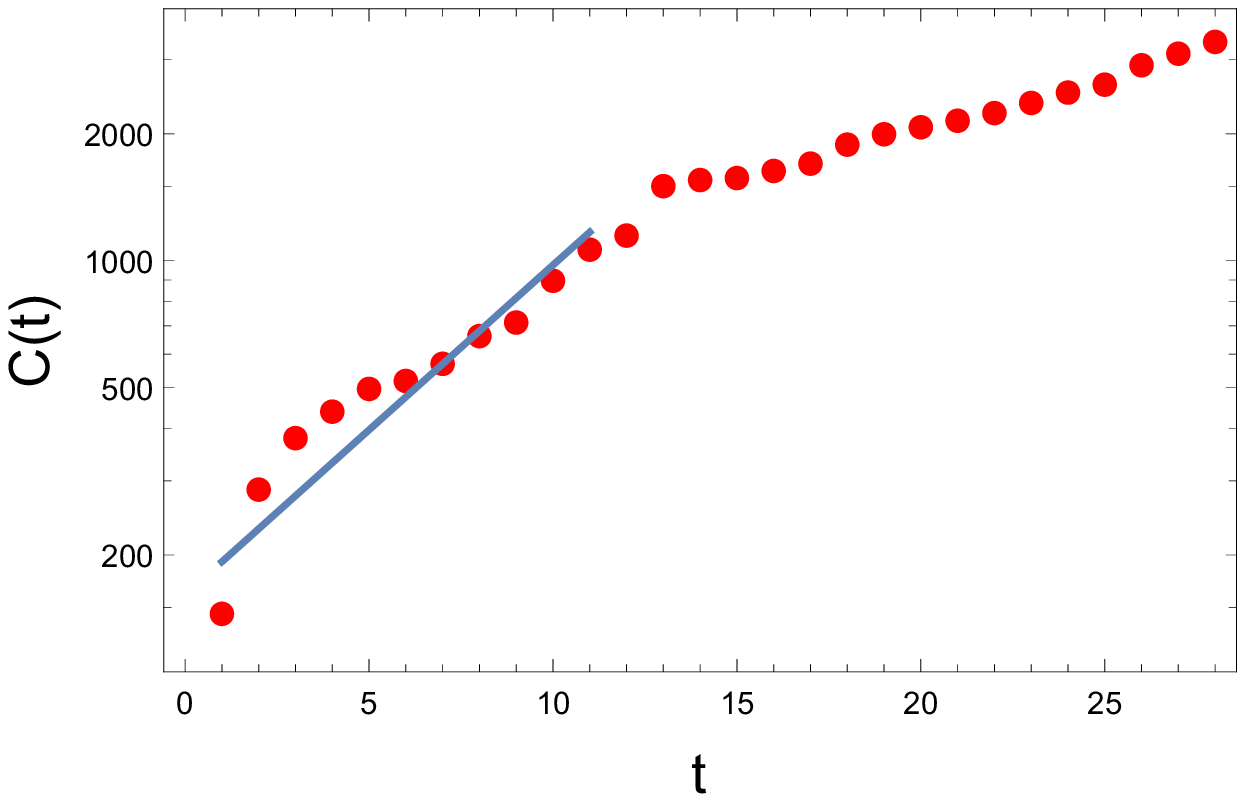}
\includegraphics[scale=0.5]{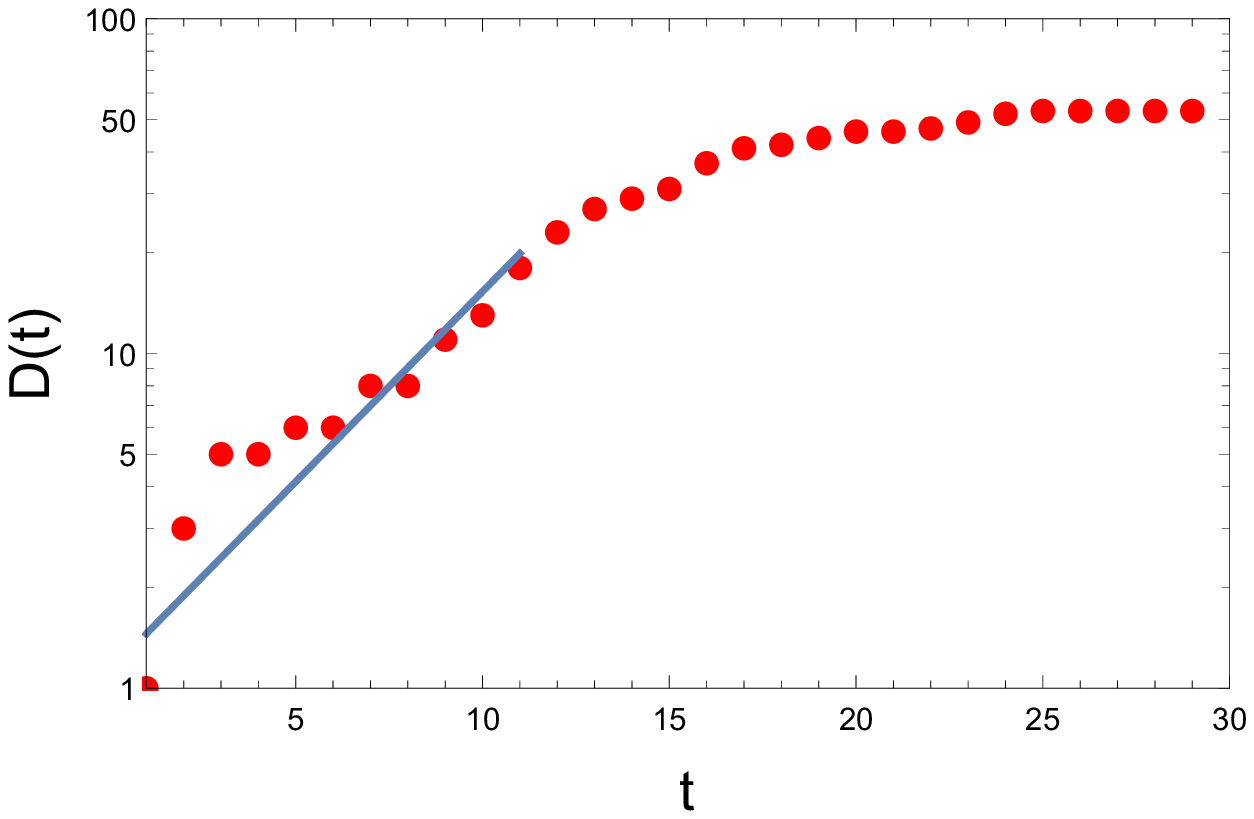}
\includegraphics[scale=0.5]{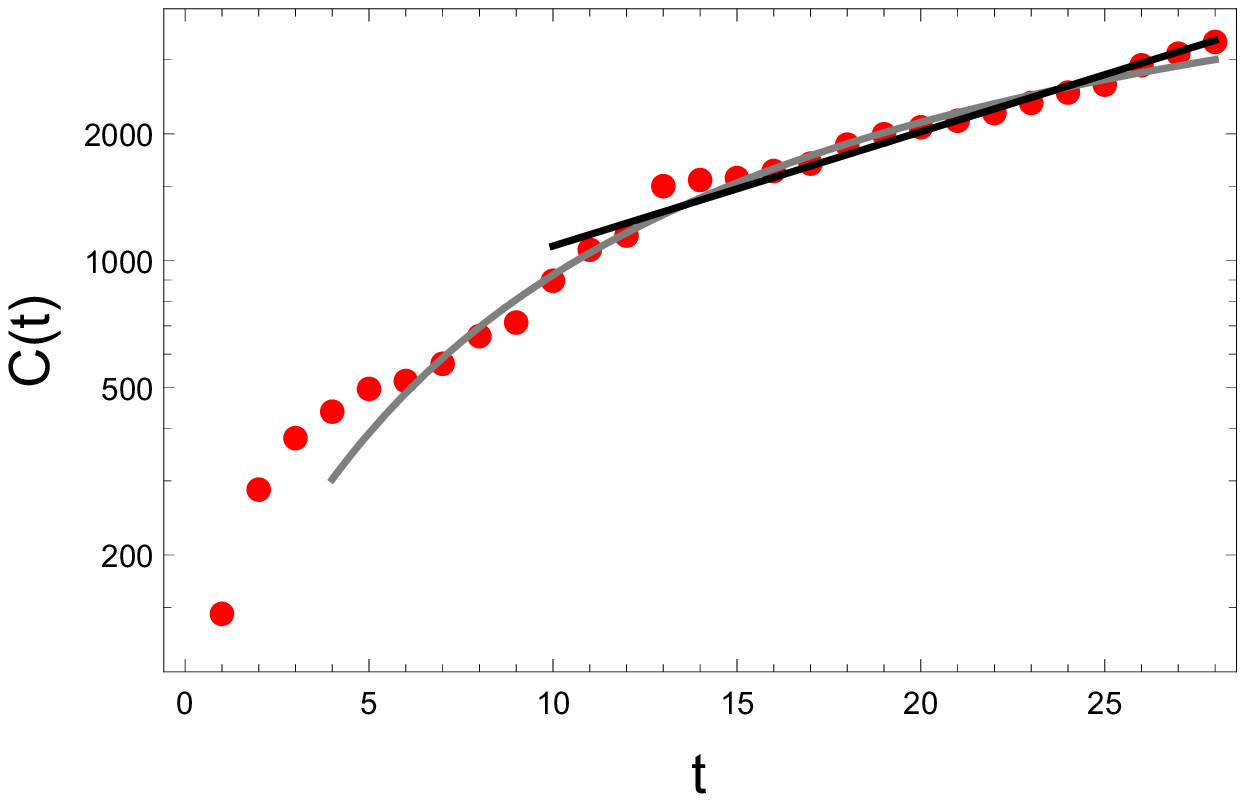}
\includegraphics[scale=0.5]{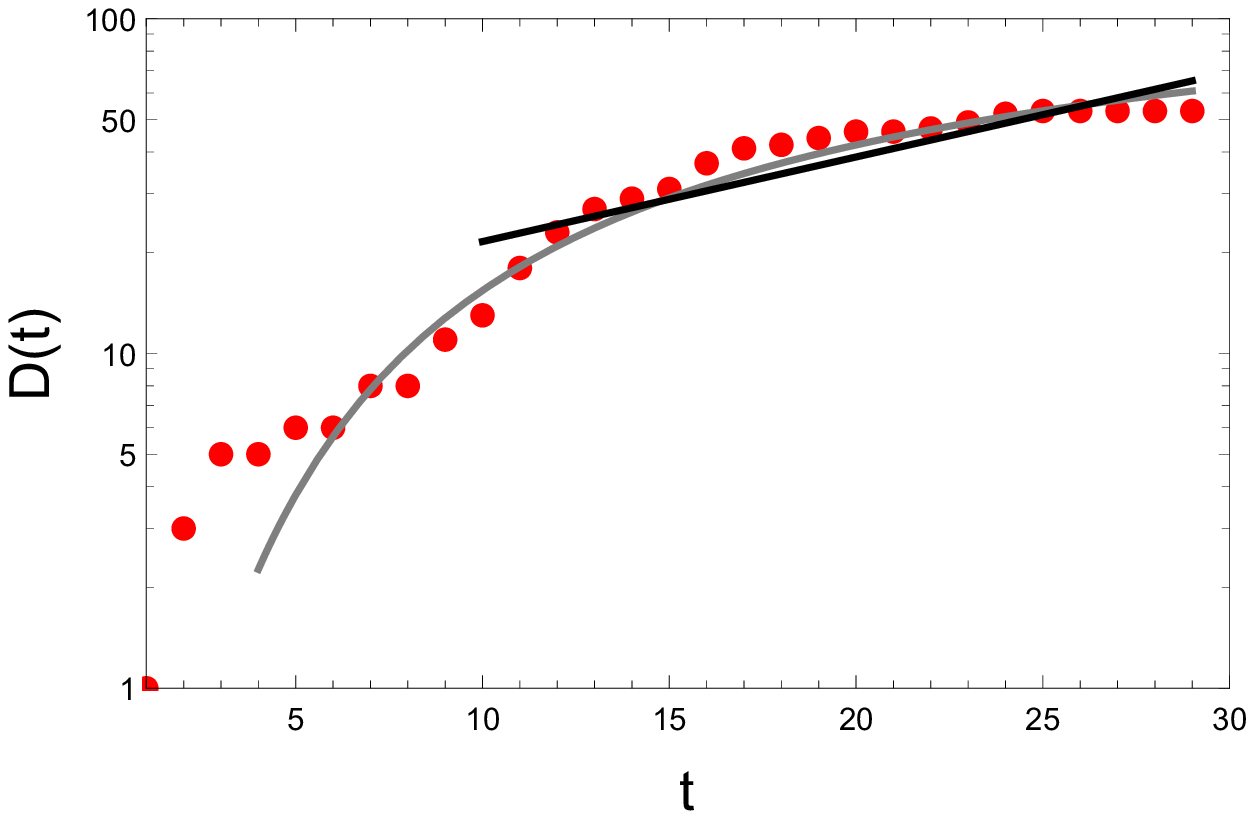}
\end{center}
\caption{The time series of $C(t)$ and $D(t)$ for Delhi shown in
 logarithmic plot; the red dots show the data. The panels at the top
 show that both sets of data indicate a change of slope around day 11,
 \ie, about 18 days after the beginning of the national lock-down. This
 requires explanation. The blue lines in these panels show the most likely
 exponential models. The lower panels show the most likely model with
 changing doubling interval (gray curves) and with a constant doubling
 interval after day 11 (black lines).}
\eef{timedep}

\bet
\begin{center}
\begin{tabular}{l|c||c|c||c|c}
\hline
Urb & Dates & $\tau_0$ & $\tau_1$ & $R_0^0$ & $-R_0^1/T$ \\ \hline
Ahmedabad & 08/4--28/4 & 1.63 & 0.17 & 8.6 & 1.6 \\
Chennai & 03/4--19/4 & 4.94 & 0.19 & 3.5 & 0.2 \\
Delhi & 31/3--28/4 & 1.12 & 0.11 & 12.0 & 2.2 \\
Indore & 31/3--28/4 & 1.06 & 0.13 & 12.6 & 2.9 \\
Mumbai & 31/3--28/4 & 1.64 & 0.13 & 8.5 & 1.2 \\ \hline
Pune & 04/4--28/4 & 0.58 & 0.12 & 22.3 & 8.8 \\
\hline
\end{tabular}
\end{center}
\caption{Parameters of the most likely model describing the growth of
 fatalities in several urban regions (cities, except for the district
 of Pune). The most likely initial doubling interval $\tau_0$ and the
 linear coefficient of change $\tau_1$, along with estimates of the
 initial $R_0^0$ and its rate of change $R_0^1$.  Note the extremely
 large values of $R_0^0$.}
\eet{r0cities}

The time series of $C(t)$ and $D(t)$ is shown for the example of Delhi
in \fgn{timedep}. Of the regions that we analysed, most cities show an
initial rapid growth followed by a tempered growth. The exceptions are
Ahmedabad and Chennai among cities, and the states of Gujarat, Kerala,
and West Bengal.  Note that day one is taken to be March 31, 2020, which
is 7 days after the beginning of the national lock-down. Since $t_2=
17.8$ days, it might be expected that the growth rate of cases in the
pre-lock-down period could manifest itself in that of fatalities until
around day 11. In case of a successful lock-down, $D(t)$ could then show
an initial exponential growth, tempered after day 11.  The initial data
for fatalities in Delhi, Indore, Mumbai and Pune can indeed be described
by an exponential. However, the doubling interval in Pune turns out to
be half of that in Mumbai, although the average population density of
Mumbai is about 6 times larger than the average in Pune city. 

The ansatz of \eqn{timedependent}, \ie, a linearly varying doubling
interval, was also examined for urban regions. The results are collected
in \tbn{r0cities}. In most locations the initial doubling interval seems
to be between half and day and two days. When converted to $R_0$, one
obtains extremely high values, far in excess of what has been quoted in
the literature. Certainly $R_0$ could vary from place to place, since
it depends on infectivity of the virus as well as the social networks
in each location, and the latter may change from one place to another.
However, $\tau_0$ for Pune is one third that of Mumbai, when Mumbai has
six times the average population density.  The wrong dependence of the
doubling time for fatalities on population density, together with the
observation that $C(t)$ shows a growth till the same date, supports the
idea that there could be a more parsimonious explanation for this common
period of growth. This is discussed in the next section. At the moment,
any statistical evidence for a gradual slowing of the growth rate of
the epidemic is hidden due to some confounding factors.

\bet
\begin{center}
\begin{tabular}{c||l|c||c|c|c|c|c|c}
\hline
Data set & City & Dates & \multicolumn{3}{c|}{$\tau$ (days)} 
                        & \multicolumn{3}{c}{$R_0$} \\ \cline{4-9}
 & & & med & IQR & \cl{95} & med & IQR & \cl{95} \\ \hline
Cases
&Ahmedabad&11/04/20&	4.32&4.24:4.42&4.11:4.55&3.86&3.81:3.92&3.71:4.01 \\
&Chennai  &10/04/20&	7.52&6.87:8.25&5.92:10.2&2.70&2.55:2.85&2.27:3.13 \\
&Delhi    &11/04/20&	11.7&10.7:13.0&9.00:17.0&2.10&1.99:2.21&1.80:2.42 \\
&Indore   &11/04/20&	7.20&6.85:7.61&6.20:8.43&2.74&2.65:2.83&2.49:2.99 \\
&Mumbai   &10/04/20&	7.35&7.00:7.72&6.45:8.47&2.70&2.62:2.78&2.46:2.94 \\
\cline{2-9}
&Pune     &10/04/20&	6.68&6.36:7.00&5.85:7.75&2.87&2.78:2.95&2.62:3.12 \\
\cline{2-9}
&Gujarat  &11/04/20&	5.30&5.06:5.55&4.68:6.09&3.35&3.24:3.46&3.05:3.65 \\
&Kerala   &31/03/20&	34.5&29.7:41.0&24.0:65.0&1.39&1.34:1.45&1.24:1.55 \\
&W.Bengal &11/04/20&	6.91&6.52:7.42&5.83:8.51&2.82&2.71:2.93&2.49:3.15 \\
\hline
Fatalities
&Ahmedabad&11/04/20&	4.85&4.56:5.15&4.15:5.78&3.59&3.44:3.74&3.15:4.03 \\
&Chennai  &10/04/20&	8.80&8.11:9.52&7.00:11.5&2.38&2.29:2.48&2.12:2.65 \\
&Delhi    &11/04/20&	12.4&11.4:13.6&9.78:16.5&2.03&1.94:2.12&1.78:2.28 \\
&Indore   &11/04/20&	13.7&12.3:15.4&10.3:20.1&1.95&1.85:2.05&1.67:2.24 \\
&Mumbai   &10/04/20&	9.60&8.89:10.4&7.78:12.2&2.32&2.22:2.42&2.04:2.61 \\
\cline{2-9}
&Pune     &10/04/20&	11.1&10.3:12.0&9.12:14.0&2.14&2.06:2.22&1.91:2.38 \\
\cline{2-9}
&Gujarat  &11/04/20&	5.11&4.89:5.37&4.50:5.89&3.44&3.32:3.55&3.12:3.66 \\
&Kerala   &31/03/20&	24.4&22.0:27.5&18.0:36.0&1.81&1.78:1.84&1.73:1.89 \\
&W.Bengal &11/04/20&	8.01&7.28:8.73&6.30:10.5&2.62&2.48:2.75&2.22:3.01 \\
\hline
\end{tabular}
\end{center}
\caption{The stable doubling interval during the lock down, $\tau$, and the
 inferred basic reproduction rate, $R_0$, for different geographical
 locations in India. Shown are the median value, the inter-quartile
 range (IQR), and the 95\% credible interval (\cl{95}).}
\eet{taufit}

In view of this, the analysis was continued with a constant doubling
interval, $\tau$, applying it to the period after day 10 or 11. For this
part of the analysis data was from three states, namely Gujarat, Kerala,
and West Bengal, was also used. From \fgn{timedep}, one sees that this
simpler model provides as good a description of the data as the model
of \eqn{timedependent}. Furthermore, this yields more realistic values
of $R_0$ implies that during the lock-down each of these places has seen
a location dependent constant doubling interval.  The values of $\tau$,
along with inferred values of $R_0$, are collected in \tbn{taufit}. These
are the primary results of this analysis.

It was noted that the number of known cases, $C(t)$, is definitely missing
cases among those who have not been tested. This could include a possibly
large, fraction of asymptomatic and non-critical or pre-symptomatic
cases \cite{asymptomatic,evacu}. However, India's disease surveillance
mechanism has concentrated on identifying critical cases and contact
tracing, which could be a good tracer of the growth of epidemics. If this
reasoning is correct, then, during the early growth of the epidemic, one
should be able to obtain reliable doubling intervals from the cumulative
counts of test positives \cite{previous}. The results of this analysis
are also given in \tbn{taufit}. The two independent estimates of $R_0$
agree well enough that a closer look reveals interesting patterns.

\bef
\begin{center}
\includegraphics[scale=0.5]{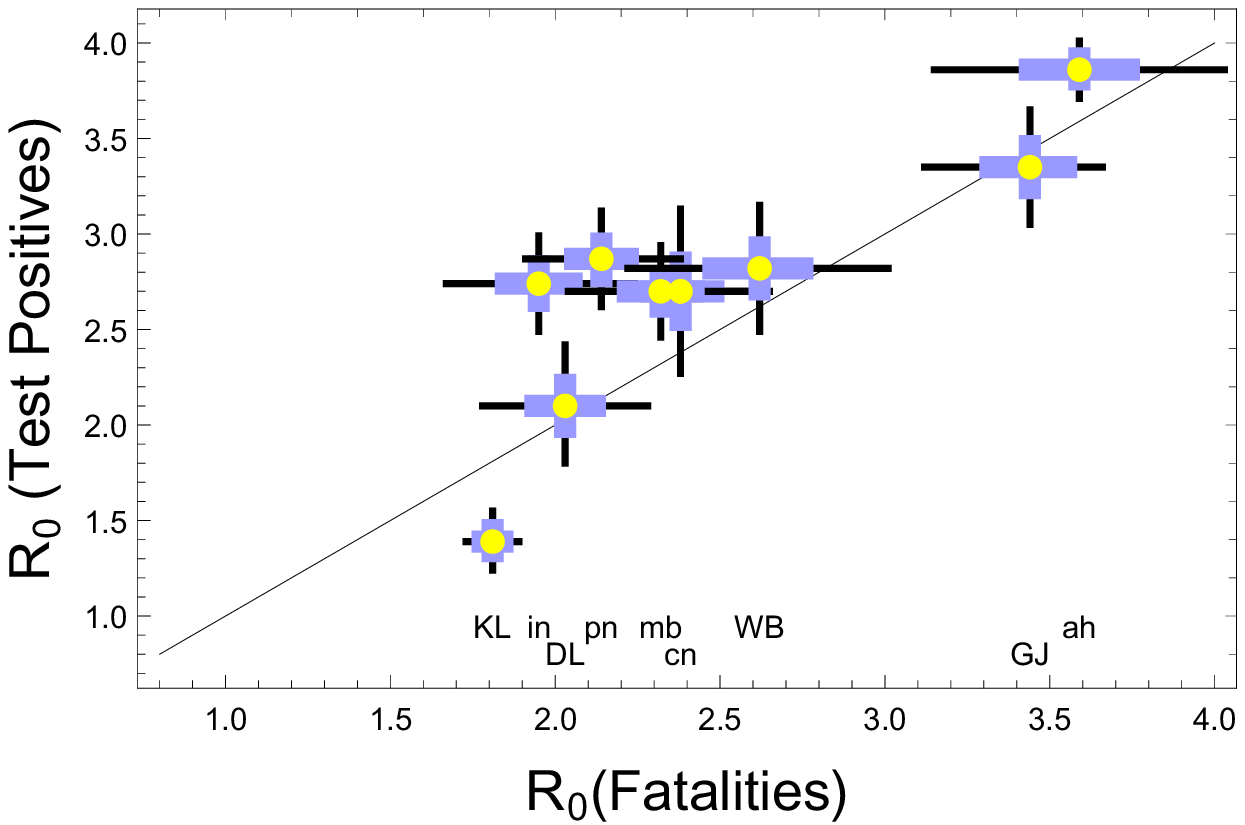}
\end{center}
\caption{The values of $R_0$ obtained from fatalities and test positives
 are plotted against each other. The diagonal (thin black) line marks the
 locus of equal values obtained from the two series. The yellow
 dots mark the medians inferred from data, the thick blue bars the IQR,
 and the black lines the \cl{95}. The code along the bottom edge identifies
 each state (DL for Delhi, KL for Kerala, WB for West Bengal) or city
 (ad for Ahmedabad, cn for Chennai, id for Indore, mb for Mumbai, and
 pn for Pune). The tag is placed at the same $R_0$ as the corresponding
 city. When two cities are at nearly overlapping, then the tags are
 displaced vertically in the same order as the results.}
\eef{results}

The scatter plot in \fgn{results} of $R_0$ obtained in two different
ways shows several interesting patterns. First, there seem to be two
groups of outbreaks. Most regions have $R_0$ below 3. Among the regions
that we studied, Ahmedabad and Gujarat were a separate group, which saw
a faster epidemic growth, with $R_0$ above 3.  Finally there is Kerala,
a different outlier, whose doubling intervals are longer than $t_2$,
and therefore with very low values of $R_0$.  The case of Kerala merits
a separate remark. The cumulative number of fatalities reached 4 at
the end of the period of study. With such low counts of fatalities the
assumption of exponential growth cannot be well tested. The counts of
total infections was larger, and supported the hypothesis of exponential
growth over the period studied.

Second, one sees that most estimates lie close to the diagonal line. If
the data was perfect, and the epidemic grew steadily, the estimates
would lie exactly on this line. With this requirement we can separate
the regions into two groups. One consisting of Ahmedabad, Chennai,
Delhi, Gujarat, and West Bengal are, within statistical uncertainties,
on this line. The second group, with Indore, Kerala, Mumbai, and Pune,
are not. This could indicate some issues with the data. 

On the other hand, if the data is as good as the other regions, then the
fact that they are off the diagonal line should be understood. Kerala,
which is the only region which lies below the diagonal, is perhaps seeing
a lower growth in new cases than fatalities, which could be indicative
of a gradual slowing down of the epidemic. Due to the lag by $t_2$,
fatalities would see the slowing down later. Conversely, the regions
which lie above the diagonal (namely Indore, Mumbai, and Pune, and,
possibly, Chennai) could be seeing an increased growth in infections,
not yet visible in fatalities because of the same time lag. Whether these
scenarios are true, or the data quality is not dependable, should be known
to the health agencies now, and would become visible to the public later.

\subsection{On CFR}

\bef
\begin{center}
\includegraphics[scale=0.5]{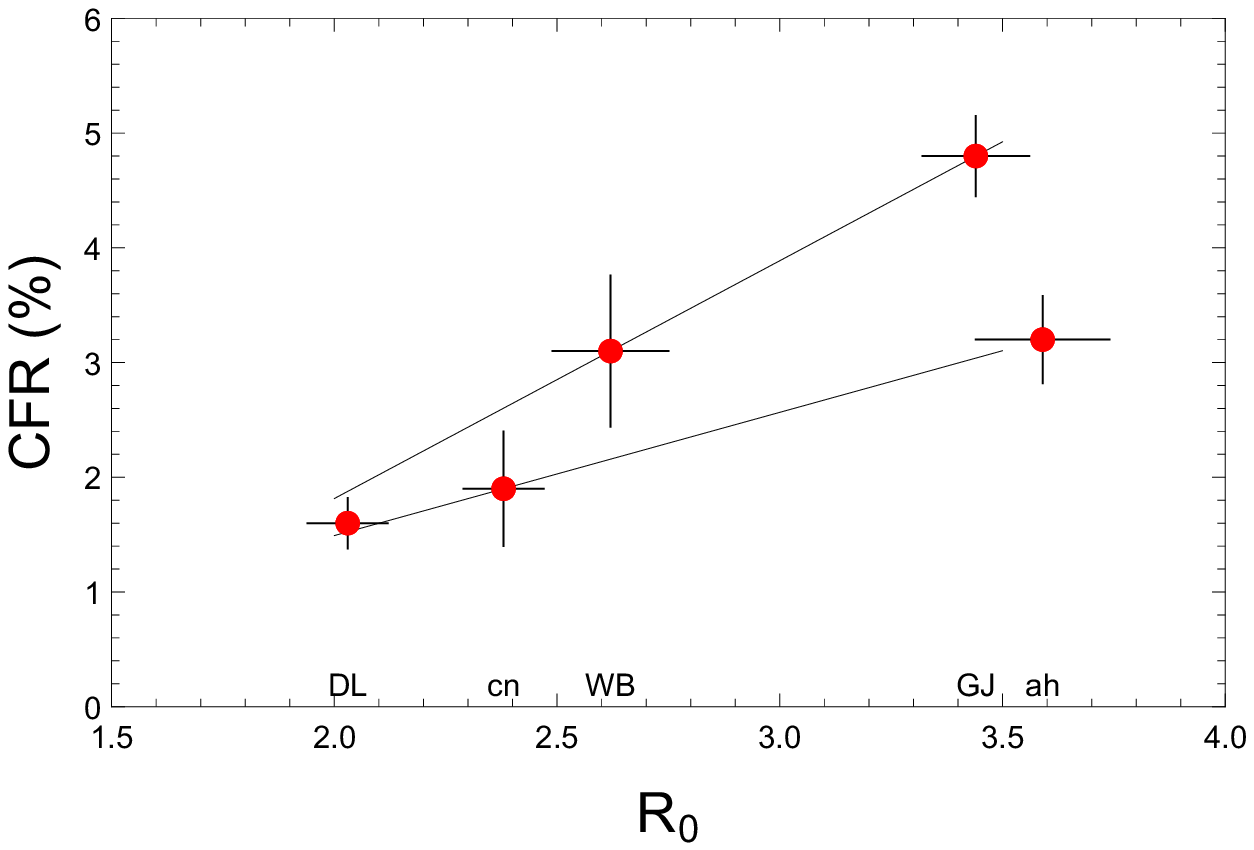}
\end{center}
\caption{The values of {\cfr} on the last day of the analysis period,
 plotted against $R_0$ obtained from fatalities.  Errors on {\cfr}
 are given by the standard deviation, and the error bars on $R_0$
 indicate the IQR. One of the trend lines joins the data for cities,
 the other for states. The tag along the bottom identifies each state
 (GJ for Gujarat, WB for West Bengal) or city (DL for Delhi, ad for
 Ahmedabad, cn for Chennai).  It is placed directly below the
 symbol for the corresponding region.}
\eef{cfrplot}

When there is a statistically significant difference between the
doubling interval determined by $C(t)$ and $D(t)$, then the ratio gives
a time-dependent \cfr. This is usually understood to be a transient
phenomenon. In view of this, the analysis was restricted to Ahmedabad,
Chennai, Delhi, Gujarat, Kerala, and West Bengal, \ie, the regions
which lie on the diagonal line of \fgn{results}, and therefore are
seeing a steady growth of identified cases as well as fatalities. The
case fatality ratios for these regions are plotted against the $R_0$
inferred from $D(t)$ in \fgn{cfrplot}.

The most obvious trend is that for the group of three cities there is
an overall trend towards smaller {\cfr} with decreasing $R_0$. This is
also true of the two states. However, the {\cfr} for states is displaced
upwards from that for cities. Both trends have strong implications for the
public health outlook and will be discussed further in the next section.

\section{Discussion}

Counts of known cases and fatalities of COVID-19 from five cities
(Ahmedabad, Chennai, Delhi, Indore, and Mumbai), one district (Pune),
and three states (Gujarat, Kerala and West Bengal) was investigated
in this work. In two of the groups, there was one case each where the
epidemic was not severe at the end of April (Chennai among cities,
and Kerala among states).  The others were known hot spots. Kerala is
special because the number of fatalities is too low for statistical
tests to be meaningful.  There are strong regional heterogeneity in
the course of the epidemic, indicating the necessity of looking at its
spread at extremely local scales in order to check and control it.

\subsection{Understanding the rapid initial growth}

The time series both of known cases and fatalities in four out of
the six urban centers showed a rapid rise for about 18 days after
the lock-down. This was followed by a much slower growth. Since
fatalities track cases with a delay of 17.8 days on the average, the
early part of this data could track the growth in the time before the
lock-down. However, it turns out that the data grows faster in less dense
urban areas. Moreover, this hypothesis is not tenable for the growth in
the number of known cases.

A possibility which resolves these difficulties is that this rapid rise
of numbers in the early days tracks the rapid improvement of disease
surveillance rather than the epidemic. The fact that the positive cases
in Kerala does not show such a rapid initial growth is consistent with
reports that the state activated disease surveillance after the first
infections came from abroad \cite{keralasurv}. This could also be true
of Ahmedabad and Gujarat, two other centers which show no such initial
increase, since the state had passed through the surveillance challenge
of Zika virus in recent years \cite{zika}.  Due to this confounding
factor, it is not possible to use the data until April 10 or 11 to make
any statistically valid measurement of the growth of the epidemic before
the lock-down. Neglecting this leads to multiple fallacies, which I remark
on next.

\subsection{Fallacies}

The apparent slowing down of the growth in later stages may be falsely
interpreted as a transition to polynomial growth. As shown in \eqn{extrap}
and \eqn{matching}, this is equivalent to a time dependent doubling
interval.  It has been discussed in the previous subsection that this
leads to highly unlikely properties of the COVID-19 epidemic.

The same apparent slowing down of the growth rate in India has also been
interpreted within the homogeneous SIR model with constant, time invariant,
parameters \cite{singapore}. In such a simple model the time dependence
can only come from early evolution towards herd immunity. This gives rise
to the unlikely conclusion that herd immunity will be reached for COVID-19
while 99\% of the population remains susceptible.

A misrepresentation of data also arises when ``instantaneous doubling
intervals'' or similar measures of exponential growth are constructed
using $C(t)$ for one day, or averaged over small windows of time
\cite{owid}. This shows a spurious gradual slowing of growth during the
first three weeks of the lock-down. In later weeks these estimates are
also plagued by spurious effects which result when delayed reports are
dumped into cumulative numbers on one day instead of being assigned to
correct past dates. These appear as evidence of local spurts or slumps
in growth. Evidence of retroactive corrections from \cite{ahmedabad}
shows that delays of as much as ten days may occur.  When these artifacts
are averaged over a moving window, this gives the mistaken appearance
of peaks and troughs, and may put erroneous pressure to change policies.

\subsection{Doubling time and $R_0$}

Due to the reasons discussed in the previous subsections, the period
after April 10 or 11 constitutes the base data for the main part of
this analysis. As shown in \fgn{timedep} a constant growth rate in each
locality during the the lock-down models the data as well as a growth
rate which changes linearly with time. This is also the most parsimonious
hypothesis about the growth of the epidemic.

The observed doubling interval, and the derived quantity $R_0$, fall into
three groups (see \tbn{taufit} and \fgn{results}). Several geographical
regions have $R_0$ less than 3. Kerala has $R_0\simeq1.7$ (a doubling
interval larger than $t_2$, the interval from the emergence of symptoms
to death). Gujarat and Ahmedabad have $R_0$ higher than 3. Since this is
the growth rate during the lock-down, population density effects are unlikely
to be the major determinant of $R_0$. It would be worthwhile to consider
the role of individuals with extremely large number of contacts in
this context, or a significant tail of the distribution with small number
of contacts, but still above three.

Five regions pass the following data quality test--- the value of $R_0$
obtained from the growth of fatalities and cases are equal. This does not
mean that the number of cases is correctly counted. Rather it indicates
that the effort to find the cases requiring critical care, and tracing
their contacts has successfully resulted in tracking a constant fraction
of all infected persons. It may miss, for example, a large fraction of
asymptomatic cases.

\subsection{Chances of fatality}

For the five geographical regions which pass the quality test described
in the previous section, a further study was performed. The dependence
of the case fatality ratio, {\cfr} (\ie, the ratio of the observed number
of fatalities and cases) on $R_0$ was investigated. Although the number
of cases identified may be much smaller than the actual number of cases,
the chance that cases are identified in these five regions are expected
to be similar, since the rate of testing is about the same. A positive
correlation between $R_0$ and {\cfr} is observed.

One possible reason for this is that with lower $R_0$ the number of
critical cases grows slower, giving medical practitioners time to figure
out good practices which prevent critical care patients from progressing
to fatality. Deeper studies of this factor, comparing case data from
different regions, is called for in future. It is possible that this is
one of the most positive, and least discussed, outcomes of the lock-down.

Another possibility may also be conjectured. Careful maintenance of social
distancing, necessary to reduce $R_0$, results in evolutionary pressure
on the virus. Lock-down and similar methods force the virus to evolve
in a direction which maximizes its ability to reproduce, which it can do
if the disease becomes less critical or asymptomatic, and the chances of
fatality decrease. It would be interesting to compare different regions
across the globe for changes in the serial time and {\cfr}.

\subsection{Public health implications}

At the observed rate of growth, and with the current rate of testing, more
than 0.5\% of the population in hot spots will begin to test positive for
infections in about a month. A constant rate of growth of infections means
that the number of hospital beds will also grow at the same rate, for as
long as the epidemic is growing. Even if the rate is slowed down heavily,
as it is already in Delhi, Mumbai, and Chennai, the demand for hospital
facilities will keep on growing, as long as the epidemic grows.

This demand is already beginning to outstrip resources in the larger
cities. The mean interval between the start of symptoms and discharge
was estimated to be 24.7 days \cite{lancet}. This means that unless the
doubling interval is kept above 35 days ($=24.7/\ln 2$), the demand on
hospitals will keep rising. Of the places we studied, only Kerala has
begun to approach this break-even point.

{\cfr} is currently small, partly because medical facilities have been able
to cope with the rate of growth. If the number of cases exceeds the capacity
of the medical system, cases which might have recovered will be harder to
treat. Inevitably in such cases {\cfr} will climb. It is useful to note that
in \fgn{cfrplot} the statewide figures for {\cfr} are higher than those for
cities. This is a reflection of the relative paucity of medical services
outside cities, and is a pointer to what might happen when the number of
infections rises beyond the sustainable capacity of hospitals.

\section{Acknowledgements}

I thank
Rahul Banerjee,
Prahlad Harsha,
D.\ Indumathi, and
R.\ Shankar
for sharing collated data on various cities.
I thank Jayasree Subramanian for providing me with the reference
\cite{ahmedabad}.

\appendix

\section{Time varying $R_0$ and doubling interval}

In this appendix the unit of time will be taken to be the inverse of
the mean rate of fatality of the infected. In these units, $R_0$ is the
average number of new infections caused by an infected person. $R_0$
depends on the infectivity of the virus, as well as an average degree
of the contact network. As a result, it may be affect by public health
policies, such as a lock down. Say a policy measure has a time scale is
$T$. Due to this, $R_0$ may become time-dependent, and one may write a
Taylor series expansion
\beq
  R_0 = R_0^0 + R_0^1\,\frac tT + \frac12 R_0^2\,\left(\frac tT\right)^2
   + \cdots
\eeq{taylor}
One may introduce this into a typical epidemic model equation, to obtain
\beq
  \frac{dI}{dt} = (R_0-1) I, \quad{\rm which\ gives}\quad
   \log\left[\frac{I(t)}{I(0)}\right] = t\left[ R_0^0-1 
      + \frac1{2!} R_0^1 \left(\frac tT\right) 
      + \frac1{3!} R_0^2 \left(\frac tT\right)^2
      + \cdots\right].
\eeq{infect}
In this form of the equation time is measured in units of the case
resolution time.  This equation assumes that the fraction of susceptible
persons is close to unity, and the fraction of persons in any other
compartment is very small. As argued before, this is a reasonable
assumption to make.  The cumulative number of infections is then found
by integration. There is no closed form result for the general case. If
only the linear term in the expansion of $R_0$ is retained, then
\beq
   \I(t) = I_0 \sqrt{\frac{\pi T}{2R_0^1}} {\rm e}^{-T(R_0^0-1)^2/(2R_0^1)}
         \erfi\left[\frac{T(R_0^0-1)+R_0^1 t}{\sqrt{2TR_0^1}}\right].
\eeq{cumulative}
The function $\erfi$ is defined through the integral
\beq
   \erfi(z) = \frac2{\sqrt\pi}\int_0^z dx\,{\rm e}^{x^2}\\
\eeq{erfi}
It is possible to use an expansion for $t\ll T$, which gives the form
\beq
   \I(t) = \I(0)\frac1{\lambda_0} {\rm e}^{\lambda_0 t} \left[
     1+\epsilon(1-\lambda_0t+\lambda_0^2 t^2) + {\cal O}(\epsilon^2)\right]
\eeq{extrap}
where the notation $\lambda_0=R_0^0 -1$, and 
$\epsilon=R_0^1/(\lambda_0^2 T)$ are introduced. The imaginary part vanishes exponentially.
This is easy to match to the phenomenological form
\beq
   \I(t)=\I(0)\,2^{t/(\tau_0+\tau_1t/T')} = \I(0)\,2^{t/\tau_0}\left[
    1-\left(\frac{\ln2\tau_1}{\tau_0^2 T'}\right) t^2
           +{\cal O}\left(\frac1{{T'}^2}\right)\right],
\eeq{phenoform}
where an artificial expansion parameter $T'$ is introduced. It is
set to unity after expansion. Matching these two expansions is accurate only
when $\lambda_0t$ is large. Then
\beq
 R_0^0 = 1+\frac{\ln2}{\tau_0}, \qquad{\rm and}\qquad
 \frac{R_0^1}T = -\frac{2\tau_1\ln2}{\tau_0^2}.
\eeq{matching}
The phenomenological parametrization of \eqn{infect} can be connected to
the parameters of (non-autonomous) evolution equations for the epidemic.
Note that $T$ and $T'$ are both regularization scales, in the sense of a
renormalization group, whose numerical value need not be specified.

In order to change units of time to days, it is necessary to choose a
model of the epidemic. If one uses the SEIR model, then the unit of time
would be the median interval between the appearance of symptoms and
the time of fatality or recovery, whichever is earlier. This quantity,
$t_2=17.8$ days \cite{lancet}. If one instead uses the SIR model,
then it is appropriate to choose the unit of time to be the median
interval between the beginning of the infection and the earlier of the
time of fatality or recovery. This is $t_1+t_2$, where $t_1$ is the median
pre-symptomatic period, $t_1=5.1$ days. Here the conversion is made
within the SEIR scheme. This gives
\beq
 R_0^0 = 1+\frac{t_2\ln2}{\tau_0}, \qquad{\rm and}\qquad
 \frac{R_0^1}T = -\frac{2t_2\tau_1\ln2}{\tau_0^2}.
\eeq{match}
When a constant $\tau$ is used, one can set $\tau_1=0$ in the above
formul{\ae} and write $R_0$ and $\tau$ instead of $R_0^0$ and $\tau_0$.

\end{document}